# LINK STREAM GRAPH FOR TEMPORAL RECOMMENDATIONS

Armel Jacques Nzekon Nzeko'o[1,2,3,*], Maurice Tchuente[2], Matthieu Latapy[3]

[1] Sorbonne Université, IRD, UMI 209 UMMISCO, F-93143, Bondy, France
[2] Université de Yaoundé I, CETIC, Faculté des Sciences, Département d'Informatique, BP 812, Yaoundé, Cameroun
[3] Sorbonne Université, CNRS, Laboratoire d'Informatique de Paris 6, LIP6, F-75005, Paris, France

## ABSTRACT

Several researches on recommender systems are based on explicit rating data, but in many real world e-commerce platforms, ratings are not always available, and in those situations, recommender systems have to deal with implicit data such as users' purchase history, browsing history and streaming history. In this context, classical bipartite user-item graphs (BIP) are widely used to compute top-N recommendations. However, these graphs have some limitations, particularly in terms of taking temporal dynamic into account. This is not good because users' preference change over time. To overcome this limit, the Session-based Temporal Graph (STG) was proposed by Xiang *et al.* to combine long- and short-term preferences in a graph-based recommender system. But in the STG, time is divided into slices and therefore considered discontinuously. This approach loses details of the real temporal dynamics of user actions. To address this challenge, we propose the Link Stream Graph (LSG) which is an extension of link stream representation proposed by Latapy *et al.* and which allows to model interactions between users and items by considering time continuously. Experiments conducted on four real world implicit datasets for temporal recommendation, with 3 evaluation metrics, show that LSG is the best in 9 out of 12 cases compared to BIP and STG which are the most used state-of-the-art recommender graphs.

## 1. INTRODUCTION

The ever-growing number of items available on e-business, booking, or news platforms makes it crucial to help users finding the best ones for them. This is the goal of recommender systems, which are generally classified in two categories: rating prediction and top-N item recommendation [1]. The goal of the first one is to predict the rating value that a user is likely to give to an unrated item. The goal of the second one is to rank all items for a giving user and propose him/her the N most interesting ones.

Many state-of-the-art recommender systems tackle the rating prediction problem [2]–[4]. However, in many on-line platforms we observe the lack of explicit rating and the abundance of data in browsing history that link users to items and conserve the timestamps of the user action. Indeed, in those platforms, recommendations are offered to users in the form of list. All this justifies the fact that top-N recommendation is becoming the most used standard of the field [5].

To achieve the goal of top-N recommendation, the most used and the most studied approach is collaborative filtering (CF) which is based on correlation between user interests [5]–[7]. Some of CF techniques are graph-based and generally built on classical bipartite graph (BIP) [8]. This graph failed to accurately capture the users' interests' dynamics. This is an important limitation because the tastes and preferences of users evolve over time. It may be due to the fact that items are out the date or because user profiles are changing.

To tackle this issue, Xiang *et al.* [9] proposed Session-based Temporal Graph (STG) which simultaneously models users' long-term and short-term preferences over time. Time is divided in slices to model users' short-term preferences in the STG. This last remark shows that STG considers time discontinuously. Thus, details of the temporal or structural dynamics of user actions can be lost depending on whether the slice size is very large or very small. So, the problem of accurately capturing the dynamics of user preferences remains.

In this paper, we address this challenge by proposing the Link Stream Graph which is an extension of link stream representation proposed by Latapy *et al.* [10] and which allows to model interactions between users and items by considering time continuously. Before describing our model in Section 4, we first present the problem statement in Section 2, following by related work on graph-based recommender systems in Section 3. Section 5 presents experiments setup and results on four real world datasets. Section 6 is dedicated to discussion and Section 7 concludes this work.

## 2. PROBLEM STATEMENT AND DATA MODELING

In this section, we present the top-N recommendation problem, for which we propose a new graph, and the data modeling formalism adopted for input data.

### 2.1. Problem statement
We follow here the time-aware top-N recommendation definition of Stefanidis *et al.* [11]: we consider an integer $N$, a set $U$ of users, a set $I$ of items, and the recommender system aims at finding a set $R_{u,t} \subseteq I$ of $N$ items that are most likely to be of interest for user $u$ at time $t$.

### 2.1. Data modeling
The recommendations rely on data regarding past interest of users in U regarding items in $I$. We model these as a bipartite link stream $L = (T, U, I, E)$ where $U$ and $I$ are the sets of users and items defined above, $T$ is the time interval during which the interests of users in $U$ regarding items in $I$ were observed, and $E \subseteq T \times U \times I$ is a set of links $(t, u, i)$ indicating that user $u$ was interested in item $i$ at time $t$. If $(t, u, v)$ is in $E$, then we say that $u$ and $v$ are linked at time $t$ in $L$, or that $(t, u, v)$ in $L$. Such a link stream typically models purchases ($u$ bought product $i$ at time $t$), interest in a cultural item (like movie watching or song listening), etc. See [12], [13] for a full description of the link stream formalism.

### 2.1. Guiding example
In the following, we will always use the following guiding example. The set of users is $U = \{u_1, u_2\}$, the set of items is $I = \{i_1, i_2, i_3, i_4\}$, the observation period is $T = [\alpha, \omega]$ where $\alpha$ and $\omega$ are constant, and $E = \{ (t_1, u_1, i_1), (t_1, u_2, i_3), (t_2, u_1, i_2), (t_2, u_2, i_3), (t_3, u_2, i_4), (t_4, u_1, i_3), (t_5, u_2, i_4), (t_6, u_1, i_2) \}$, which means for instance that $u_1$ was interested in item $i_2$ at time $t_2$. Figure 1 illustrates the link stream describe above.

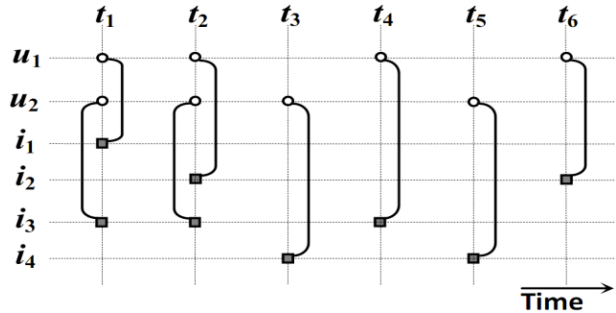

*Figure 1. Example of link stream describes in Section 2.3.*

## 3. RELATED WORK

In the following, we make a formal description of the classical bipartite graph (BIP) and its extension by the Session-based Temporal Graph (STG). This section ends with the description of personalized PageRank used to compute top-N recommendations on graphs.

### 3.1. Classical bipartite graph
The first graph we consider is the classical bipartite graph [8], that we denote by BIP. It is a directed bipartite graph $(U, I, E)$ where $U$ is the set of users, $I$ the set of items, $E \subseteq U \times I$ the set of links such that $(u, i) \in E$ if user $u$ was interested in item $i$ during the observation period, *i.e.*, there exists $t \in T$ such that $(t, u, v)$ in $L$. In addition, and to ensure consistence with other recommender graphs defined below, we consider a trivial weight function $w$ such that $w(x, y) = 1$ for all $(x, y) \in E$. See Figure 2(a) for an illustration.

We can deduce an user- or item-projection of BIP [14]. However, projected graphs lose information about user-item interactions, especially the aspect: who selected what? In addition, the update of those graphs is more expensive in dynamic contexts as time aware systems. These remarks justify the fact that the classical bipartite graph is the most used in recommender systems, particularly when temporal aspects are taken into account. Figure 2 illustrates the classical bipartite graph, the user-projected and the item-projected graphs of the link stream describes in Section 2.3.

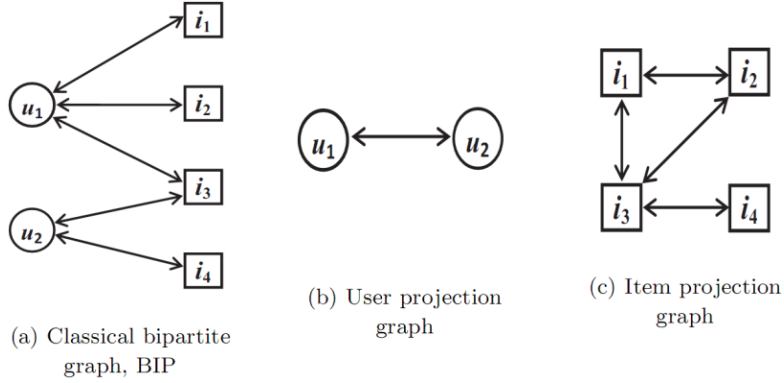

(a) Classical bipartite graph, BIP
(b) User projection graph
(c) Item projection graph

*Figure 2. Classical bipartite graph, User projection and Item projection graphs obtained from our guiding example. The weight of each edge is 1.*

### 3.2. Session-based Temporal Graph

Session-based Temporal Graph proposed by Xiang *et al.* [15] is a recommender graph designed to overcome the fact that BIP does not integrate time dimension. Indeed, STG combines short-term and long-term users' behaviors to accurately capture user's preferences over time. Thus, in addition to the sets of user and item nodes already present in BIP, a set S of session nodes is added, and each of those nodes represents the short term preferences of a user.

This graph encodes time information with session nodes defined as follows. First, for a given $\Delta$, the observation interval $T$ is divided into $|T|/\Delta$ time slices $T_k = [(k-1) \cdot \Delta, k \cdot \Delta]$ of equal duration $\Delta$. Then, S contains the couples $(u, T_k)$ such that there exists a link $(t, u, i)$ in E with $t \in T_k$. In other words, each user leads to a session node $(u, T_k) \in S$ for each time interval $T_k$ during which this user was active. This leads to the definition of STG as a tripartite graph $(U, I, S, E)$ where $U$ and $I$ are defined as for BIP, $E$ contains the links of BIP and in addition the links between $(u, T_k)$ and the items selected by user $u$ during time slice $T_k$.

In addition, we consider the directed weight function $w$ such that, for a given constant $\eta_s$, for all $(x, y)$ in $E$, $w(x, y) = \eta_s$ if $x \in I$ and $y \in S$, $w(x, y) = 1$ otherwise. See Figure 3 for an illustration with $\Delta = 3$, leading to two time slices $T_1$ and $T_2$ and three session nodes.

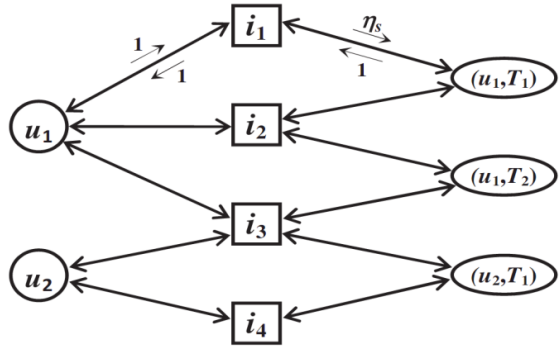

*Figure 3. Session-based temporal graph (STG) for the guiding example.*

### 3.3. Recommendation on graphs

In the previous section, we present some recommender graphs; we now show how to use them to actually solve the top-N recommendation problem. Once we have a graph, it is necessary to apply an algorithm to get the top-N recommendation. Most of these algorithms rely on the random walk, for example we have PageRank [16] and Injected Preference Fusion (IPF) [15], but also other algorithms like Hypertext-Induced Topic Selection (HITS) can be used [17].

In this paper, we focus on PageRank because it is very efficient and widely used in recommender systems [18], [19]. More precisely, we use the Temporal Personalized Random Walk (TPRW) defined by Xiang *et al.* [15], which is a personalized extension of the classical PageRank [16]. It is defined according to the idea of Haveliwala *et al*. [20] using the following equation:

$$PR = \alpha \cdot M \cdot PR + (1 - \alpha) \cdot d$$

where α is the damping factor, $M$ is the transition matrix of the graph and $d$ is a user-specific personalized vector indicating which nodes the random walker will jump to after a restart.

To recommend new items to user $u$ at time $t$, vector $d$ assumes that users tend to jump to their own user nodes and session nodes after a restart. We obtain for a given user $u$ in BIP: $d(u) = 1$ and $d(v) = 0$ if $v \neq u$; in STG: $d(u) = \beta$, $d(u, T_k) = 1 - \beta$ if $T_k$ is $u$ most recent session node, and $d(v) = 0$ for any other node $v$. Then, we run TPRW over the recommender graph to get the preferences of user $u$ for each item $i$. The output is the set of $N$ items with higher preference.

## 4. LINK STREAM GRAPH

STG takes time dimension into account but it divides time into slices. However, slicing time loses details about dynamics of events. If slices are very short, structural aspects are lost and if very long, details on dynamics are lost. To overcome this limit, we are inspired by the link stream representation of Latapy *et al*. [10], [13] and we design the bipartite link stream graph (LSG), which considers time in a continuous way.

### 4.1. Model description

This graph is first defined by a set of nodes representing users and items over time: $\{(t, u) : \exists i, (t, u, i) \in E\}$ ∪ $\{(t, i) : \exists u, (t, u, i) \in E\}$. In other words, each user $u$ is represented by the nodes $(t, u)$ such that a link involves $u$ in $L$ a time $t$, and each item is represented similarly. We then define the set of links $\{((t, u), (t, i)) : (t, u, i) \in E\}$ ∪ $\{((t, u), (t', u)) : \exists i, (t, u, i) \in E, t' = min\{x : x > t \text{ and } \exists i', (x, u, i') \in E\} \cup \{((t, i), (t', i)) : \exists u, (t, u, i) \in E, t' = min\{x : x > t \text{ and } \exists u', (x, u', i) \in E\}$. In other words, each user node $(t, u)$ is linked to both the item nodes $(t, i)$ such that $(t, u, i) \in E$ and to the next user node representing u. Item nodes are linked similarly.

In addition, we consider the directed weight function $w$ such that, for a given constant $\eta_s$, for all $((t, x), (t', y)) \in E$, $w((t, x), (t', y)) = \eta_s$ if $x = y$ and $t > t'$, and $w((t, x), (t', y)) = 1$ otherwise. See Figure 4 for an illustration on our guiding example.

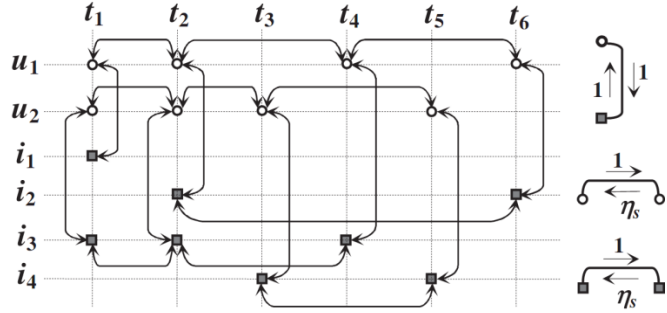

*Figure 4. Link Stream Graph for the guiding example in Section 2.3.*

### 4.2. Recommendation on Link Stream Graph

The LSG recommendation process follows the same principle as for BIP and STG in section [16]. To recommend new items to user $u$ at time $t$, vector d is personalized as follow: $d(t_k, u) = 1$ if $t_k$ is the largest value such that $t_k \leq t$ and node $(t_k, u)$ exists, and $d(v) = 0$ for any other node $v$. After running TPRW on LSG, the preference of $u$ for an item $i$ is the sum of preferences for nodes $(t_k, i)$ for all $t_k$.

## 5. EXPERIMENTS

We performed experiments to compare the performances of LSG, BIP and STG. The data used are those from 4 real-world platforms. In the rest of this section, we present datasets, experiment protocol and obtained results.

## 5.1. Data description

We use publicly available datasets, the first one is extracted from product reviews Ciao[1][21], the second one is extracted from on-line shopping website Ponpare[2]. We also used data from Delicious[3] and CiteUlike[4], which are social bookmarking websites.

Data are modeled as link stream $\{(t, u, i)\}$ and the interpretation of any triplet depends on the domain. In the case of on-line shopping, each triplet means that the customer u has bought item i at time t. In social bookmarking data, this means that user u has bookmarked page i at time t. In Ciao, we have set of tuples $(t, u, i, r)$ meaning that user u has assigned the rating $r \in \{0, 1, 2, 3, 4, 5\}$ to item i at time t. Since we work only with positive links between users and items, we discard all tuples such that the rating it contains is lower than 2.5 or the average rating of involved user.

Data are filtered such that only users and items that appear respectively at least $\sigma_u$ and $\sigma_i$ times are considered. Table 1 provides details on data used: date of the first link, date of the last link, total duration of link stream, user-threshold, item-threshold, numbers of users, items, content features and links.

*Table 1. Data statistics.*

|  | Ciao | Ponpare | Delicious | CiteUlike |
|---|---|---|---|---|
| Start date | 2007-01-01 | 2011-07-01 | 2010-05-10 | 2010-01-01 |
| End date | 2010-12-31 | 2011-11-01 | 2010-11-09 | 2010-09-01 |
| $\sigma_u$ | 1 | 10 | 4 | 10 |
| $\sigma_i$ | 1 | 10 | 4 | 10 |
| Users | 879 | 1 322 | 1 175 | 1 726 |
| Items | 6 005 | 1 115 | 1 352 | 619 |
| User-Item | 8 109 | 12 863 | 7 652 | 9 360 |
| Sparsity | 99.85% | 99.13% | 99.52% | 99.12% |
| Links | 8 109 | 177 005 | 35 558 | 24 772 |

## 5.2. Experiment and Evaluation

Evaluating recommender systems is a difficult task. In this paper, we use three classical metrics for top-N recommendations: F1-score (F1), Hit Ratio (HR) and Mean Average Precision (MAP) [22]. Higher values of these metrics indicate better recommendation performance.

F1-score is a trade-off between ranking precision and recall such that optimizing F1-score is more robust than optimizing precision or recall. Precision is the fraction of good recommendations over all recommended items and recall is the fraction of good recommendations over all relevant items to recommend. For one user $u$, $Precision = \frac{hit_N(u)}{N}$, $Recall = \frac{hit_N(u)}{I_{new}(u)}$ and $F1 = 2 \cdot \frac{Precision \times Recall}{Precision + Recall} = \frac{2 \cdot hit_N(u)}{I_{new}(u) + N}$ where $N$ is the length of recommendation list, $hit_N(u)$ denotes the number of good recommendations to $u$ in the top-N items and $I_{new}(u)$ is the set of new items to recommend to $u$. For all users the equation of F1-score is: $F@N = \frac{\sum_{u \in U} 2 \cdot hit_N(u)}{\sum_{u \in U} (I_{new}(u) + N)}$.

Hit Ratio is the fraction of users to whom the recommender system has made at least one good recommendation over all users: $H@N = \frac{\sum_{u \in U} (hit_N(u) > 0)}{|U|}$.

Mean Average Precision considers the order of items in the top-N recommendation in order to give better evaluation scores to results that recommend better items first: $M@N = \frac{\sum_{u \in U} AP_N(u)}{|U|}$ where $AP_N(u) = \frac{1}{hit_N(u)} \sum_{k=1}^{N} \frac{hit_k(u) \times h(k)}{k}$ is the average precision of top-N recommendations done to user u and $h(k) = 1$ if the $k$-th recommended item is a good recommendation and 0 otherwise.

These metrics evaluate a given top-N recommendation. Since we actually can't perform recommendations on live users, we perform evaluation on past data described above. Following a periodically evaluation process established by previous works [7], [23], [24]. For each dataset, the input link stream is divided into 8 time windows of equal length. For each time window $W_k$, for $k = 1 .. 7$, we proceed as follow:

---

[1] https://www.cse.msu.edu/~tangjili/trust.html
[2] https://www.kaggle.com/c/coupon-purchase-prediction
[3] https://grouplens.org/datasets/hetrec-2011/
[4] http://www.citeulike.org/faq/data.adp

- Build graphs that correspond to data of $W_1, W_2, .., W_k$ (training set)
- Compute Top-N recommendations for users who have selected at least one new item during the time window $W_{k+1}$ (test set)
- Compute for each evaluation metric $M$ the numerator $M_{num_k}$ and the denominator $M_{deno_k}$ of its definition, given above.

After determining $M_{num_k}$ and $M_{deno_k}$ of each time window, we compute the Time Averaged (TA) value of the concerned evaluation metric: $TA(M) = \frac{\Sigma_k M_{num_k}}{\Sigma_k M_{deno_k}}$. This leads to a time-averaged value of F1-score, Hit ratio and MAP, which we all use for evaluation. Indeed, evaluation metrics can be in disagreement [25], and so using several metrics is essential to obtain accurate insight on result quality.

### 5.3. Optimal tuning parameter estimation

There are several metrics to evaluate these graphs and some of them have more than one parameter. So, exhaustive search for the best values is out of reach. Many subtle techniques exist to explore the parameter space in search for good values. Since this search is not the focus of this paper, we use a simple approach called Randomized Search Cross-Validation [26]. This method randomly selects parameter values in a predefined set of possible values, usually designed to span well the whole set of values. Here, we use 50 such random settings, sampled in the set defined by Table 2.

*Table 2. Predefined values of parameters.*

| | parameter meaning | predefined values |
|---|---|---|
| $\Delta$ | STG session duration | 7, 30, 60, 90, 180, 365, 540, 730 days |
| $\beta$ | STG long-term preference | 0.1, 0.3, 0.5, 0.7, 0.9 |
| $\eta_s$ | weight connected to the past | 0, 0.1, 0.2, 0.5, 1, 2, 5, 10 |
| $\alpha$ | damping factor for PageRank | 0.05, 0.1, 0.15, 0.3, 0.5, 0.7, 0.9 |

### 5.4. Accuracy Comparison

Table 3 shows the best results for each recommender graph. We have dataset names in the rows, graphs and evaluation metrics used in columns. The values in bold are the best performances obtained. This table shows the relevance of the LSG which is the best 9 out of 12 times.

*Table 3. Best recommender graphs results.*

| | F1-score (%) | | | Hit ratio (%) | | | MAP (%) | | |
|---|---|---|---|---|---|---|---|---|---|
| | BIP | STG | LSG | BIP | STG | LSG | BIP | STG | LSG |
| Ciao | 1.78 | 1.73 | **3.0** | 5.63 | **6.35** | **6.35** | 2.18 | **2.23** | 2.11 |
| Ponpare | 5.52 | 6.32 | **6.67** | 11.8 | 12.4 | **29.5** | 4.03 | 4.29 | **11.7** |
| Delicious | 5.08 | 5.13 | **6.56** | 10.7 | 11.7 | **13.7** | 5.15 | 5.49 | **5.96** |
| CiteUlike | **7.65** | 7.39 | 7.48 | 28.7 | **29.5** | 28.5 | 10.7 | **10.9** | **10.9** |

Regarding the analysis of the LSG parameters, we found that the parameter $\eta_s$ is generally less than 0.5 for all datasets of this study. In the particular case of Ponpare, $\eta_s$ is always nil for the best performance. This observation confirms the importance of concept drift [4], [27] which suggest the penalization of old data to achieve good qualities of recommendation.

## 6. DISCUSSION AND PERSPECTIVES

LSG has a high space complexity, because for each link $(t, u, i)$, two nodes and three bidirectional edges are systematically created. However, in BIP (resp. STG), two nodes (resp. three nodes) and one (resp. two) bidirectional edges if they do not already exist in the graph. This complexity is more remarkable when each user selects the same product several times. Thus, a future work on LSG can be the reduction of the space complexity while maintaining or improving its performance.

In this paper, the LSG is used for top-N recommendation, but it can also be applied for link prediction problem or for any other problem with temporal and/or structural features. In such cases, other algorithms to compute prediction may be proposed, and other way of weighting edges can be proposed to optimize the LSG performance according to the problem studied.

## 7. CONCLUSION

This paper presents the Link Stream Graph (LSG) which takes time into account in a continuous way. We used it to capture temporal dynamic of users' preferences in a top-N recommender system. Comparisons with classical bipartite graph (BIP) and Session-based Temporal Graph (STG), which are the main state-of-the-art recommender graphs, using 3 evaluation metrics, confirm the relevance of the LSG which is the best 9 out of 12 possible cases. For example, there is an improvement from: 1.73 to 3.0% of the F1-score in the Ciao dataset; 11.7 to 13.7% of the Hit ration in the Delicious dataset; and 4.29 to 11.7% of the MAP in the Ponpare dataset. As future work, someone can focus on reducing the space complexity of the LSG or propose other way of weighting edges while maintaining or improving its performance.

## 8. ACKNOWLEDGEMENTS

This work is funded in part by the African Center of Excellence in Information and Communication Technologies (CETIC), the Sorbonne University-IRD PDI program, and by the ANR (French National Agency of Research) under grant ANR-15-CE38-0001 (AlgoDiv).